 \documentclass[prd,nofootinbib,showpacs,twocolumn]{revtex4-1}

\usepackage{graphicx}
\usepackage{amsmath,amssymb}
\usepackage{array}
\usepackage{subcaption}
\usepackage[colorlinks]{hyperref}
\usepackage[usenames,dvipsnames]{color}

\def\bea{\begin{eqnarray}}
\def\eea{\end{eqnarray}}
\def\bal{\begin{align}}
\def\eal{\end{align}}
 \def\be{\begin{equation}}
\def\ee{\end{equation}}
\newcommand{\VEV}[1]{\langle #1 \rangle}

\newcommand{\mgut}{M_{GUT}}

\begin{document}

\title{Topological pseudodefects of a supersymmetric $SO(10)$ model and cosmology}
\author{Ila Garg}
\email{ila.garg@iitb.ac.in}
\author{Urjit A. Yajnik}
\email{yajnik@iitb.ac.in}
\affiliation{Department of Physics, Indian Institute of Technology Bombay, Mumbai 400076, India}

\begin{abstract}
Obtaining realistic supersymmetry preserving vacua  in the minimal renormalizable supersymmetric 
$Spin(10)$ GUT model introduces considerations of the non-trivial topology of the vacuum manifold.
The $D$-parity of low energy unification schemes gets lifted to a one-parameter subgroup $U(1)_D$ of $Spin(10)$.
Yet, the choice of the fields signalling spontaneous symmetry breaking leads to disconnected 
subsets in the vacuum manifold related by the $D$-parity. The resulting domain walls, existing 
due to topological reasons but not stable, are identified as topological pseudodefects. 
We obtain a class of one-parameter paths connecting $D$-parity flipped vacua and compute the 
energy barrier height along the same. We consider the various patterns of symmetry breaking which 
can result in either intermediate scale gauge groups or a supersymmetric extension of the 
Standard Model. If the onset of inflation is subsequent to GUT breaking, as could happen also if 
inflation is naturally explained by the same GUT, the 
existence of such pseudodefects can leave signatures in the CMB. Specifically, this could have an 
impact on 
the scale invariance of the CMB fluctuations and LSS data at the largest scale.

\end{abstract}

\maketitle


\section{Introduction}
\label{sec:intro}
There are several indications for physics beyond the Standard Model (SM) which demands the need to 
connect with the high energy scales unlikely to be accessible to accelerators. One of them is the 
minuscule masses of neutrinos \cite{Hirata:1989zj,Fukuda:2001nj} which through see-saw mechanism 
\cite{Minkowski:1977sc,Yanagida:1979as,GellMann:1980vs,Mohapatra:1979ia} suggest the existence of a high mass
scale. Further the precarious hierarchy of the Higgs mass with respect to the Planck scale is
conceptually unnatural and can be easily ameliorated by new physics beyond the SM. Finally, 
unification of couplings remains ever a desirable feature and can be accomplished by grand 
unification in supersymmetric or non supersymmetric $SO(10)$, or larger gauge groups. The scale of 
such models is beyond the reach of accelerators but the early cosmology and its imprints on the CMB 
data and Large Scale Structure data can be important checks on this model. Other than the CMB, 
consistent Big Bang Nucleosynthesis  and successful inflation remain important requirements on any 
model with new physics at high energies. Indeed for the class of models that unifies, the high 
scale physics natural to them is constrained by inflation, by the need to generate baryon asymmetry, 
and by exotic relics such as cosmic strings and domain walls (DW) that survive the homogenizing 
effects of the high temperature. 

An early study \cite{Kibble:1982dd} considered the consequence of unstable domain wall formation in
$Spin(10)$, which can decay due to the formation of cosmic strings as punctures or boundaries. 
Several other 
works have considered these issues, notably \cite{Stern:1985bg,Pijush:1992, Davis:1995bx,Jeannerot:2003qv, 
Chakrabortty:2017mgi} for the context 
of topological defects and 
others 
\cite{Aulakh:2002zr,Li:2009sw,Gogoladze:2009bd,Gogoladze:2011db,LalAwasthi:2011aa,Bertolini:2012im,
Mambrini:2015vna}
which have utilized the group theoretic constraints arising from such considerations for unification 
proposals.
The present investigation is concerned with studying the interplay of symmetry breaking patterns 
with cosmology in the context of 
supersymmetric $Spin(10)$. There are two broad directions that have been pursued along these ones.
One, that is motivated by superstring unification, as in 
\cite{Blazek:2001sb,Blazek:2002ta,Ross:2002fb,Dermisek:2006dc}, and the other  
class of models relies on the minimal representations of the Higgs and has been
explored in \cite{Aulakh:1982sw,Babu:1992ia,Aulakh:2003kg,Bajc:2004xe,Bajc:2004fj,Aulakh:2008sn}. 
It has been
advanced as a renormalizable minimal supersymmetric $SO(10)$  grand unified theory (MSGUT). In the present work we shall
restrict to the latter class of models due to their rich topological structure.
The model utilizes Higgs supermultiplets {\bf10}, {\bf210}, {\bf126} ($\overline{{\bf126}}$)
 required to break the symmetry and provide fermion masses.
 Among these the ${\bf210}$ and ${\bf 126}$ ($\overline{{\bf126}}$) are responsible for breaking of
$SO(10)$ symmetry down to MSSM, and ${\bf10}$ and  $\overline{{\bf126}}$ give masses to the 
fermions. The ${\bf16}$-dim Spinor representation contains one generation of SM fermions and a right handed neutrino.  
 The GUT model 
considered has many phenomenological as well as cosmological 
virtues. It can provide inflaton candidates \cite{Aulakh:2012st,Garg:2015mra}, mechanism for 
baryogenesis through leptogenesis 
and the  lightest supersymmetric particle (LSP) of the model can serve as weakly 
interacting cold dark matter candidate.

A hallmark of this class of models is the occurrence of  $D$-parity, a discrete symmetry that can 
exchange the 
chiral matter
fields with their charge conjugates and also appropriately the corresponding Higgs bosons which 
gives them masses. While this is a discrete symmetry of the partial unification model, viz., the 
left-right symmetric model, when lifted to $SO(10)$ it gets embedded in a one parameter $U(1)_D$ 
subgroup of the covering group $Spin(10)$. Since the parent group is simply 
connected there are no stable domain walls. However we argue here that the topology of the vacuum 
manifold can be non-trivial, and can give rise to defects that are best dubbed topological 
psuedodefects (TPD). In this paper we want to study the implications to cosmology of TPD's
arising in a SUSY $SO(10)$  GUT model. In \cite{Aulakh:2003kg,Bajc:2004xe,Bajc:2004fj,Aulakh:2008sn} 
the implicit assumption is that there is a one step breaking from SUSY $SO(10)$ to MSSM.
However the physics of Big Bang implies the existence of many causally 
disconnected regions in space and the non-trivial vacuum structure  
would  give rise to  domain walls \cite{Kibble:1982dd}.

Generically, the presence of domain walls (DW) in a model has important interplay with inflation. One of the successes of 
inflationary
proposal is the removal unpleasant relics of GUTs such as monopoles. The same applies to cosmic 
strings, 
whose density can be easily diluted by cosmological inflation. The same is however not true of DW, 
as these
may form a mutually locked structure which may not be blown apart by inflation easily. This is a 
relevant 
possibility if inflation has a preceding hot period allowing DW forming phase transition. The 
walls if stable would conflict with standard cosmology due to the inhomogeneities they would 
introduce in the CMB. On the other hand,
for a network of walls which is unstable, the question shifts to the time scale of the 
decay and disappearance of the walls.
The unified model in this case can be constrained by requiring that the inhomogeneities generated 
by 
their early presence should not affect the successful outcomes of inflation, specifically the nearly 
scale
invariant CMB spectrum and the observed Gaussian nature of the density perturbations.

The plan of the paper is as follows. In section \ref{sec:KLSreview} we review the topological role 
of $D$-parity as
first considered in \cite{Kibble:1982dd}. In  \ref{sec:LRDparity}, we treat a warm up example of the 
energy barrier 
of DW in the  minimal supersymmetric left-right where $D$ parity occurs as a discrete symmetry, 
avoiding the
subtleties of the large group $SO(10)$. In section \ref{sec:DWso10}, we give a brief introduction to
the GUT model 
of \cite{Bajc:2004fj} and calculate the energy barrier associated with 
the D-parity breaking.  The height of such a barrier estimates the energy per unit area of the TPD walls that may form.
In sec. \ref{sec:DWinfl} we discuss the implications to cosmological 
inflation which in turn may imply constraints
on the scales the possible symmetry breaking schemes. The conclusions are in Sec. \ref{sec:concl}.  

\section{Metastable domain walls}
\label{sec:KLSreview}
Here we briefly review the topological issues relevant to the domain walls, paraphrased from 
\cite{Kibble:1982dd}. $Spin(10)$ can be broken to its subgroup $H_0=Spin(6)\otimes 
Spin(4)$, where the first factor contains the color $SU(3)_c$ and the second factor contains the SM 
$SU(2)_L$ and a potential $SU(2)_R$. This breaking can be achieved by using the $54$ dimensional 
scalar $\chi$ which takes on a vacuum expectation value (VEV) 
\be
\VEV{\chi} = \chi_0 \textrm{diag}(2,2,2,2,2,2,-3,-3,-3,-3).
\ee
However, the stability group of this VEV contains a discrete set of additional elements. Consider
the one-parameter curve in $Spin(10)$ of the form $U_{J_{67}}(\theta)=\exp(i \theta J_{67})$. The 6-7 
submatrix of the VEV of $\chi$ transforms under this as
\be
U_{J_{67}}(\theta)\begin{pmatrix}
2  &  0\\
0  & -3
\end{pmatrix}U_{J_{67}}^{-1}(\theta)=
\begin{pmatrix}
-\frac{1}{2}+\frac{5}{2}\cos 2\theta  &  -\frac{5}{2}\sin 2\theta\\
-\frac{5}{2}\sin 2\theta  & -\frac{1}{2}-\frac{5}{2}\cos 2\theta
                                  \end{pmatrix}.
\label{eq:fiftyfourvev}
\ee
Thus $\VEV{\chi}$ is left invariant by $U_{J_{67}}(\theta=n\pi)$ with $n\in\mathbb{Z}$. It can be seen 
that all such choices derived from mixed $a$ and $\alpha$ indices, with $a \in {1,2,...6}$ and 
$\alpha \in {7,8,9,10}$,  have the same property, and are indeed equivalent to each other under a 
suitable transformation by an element from $H_0$.  
Thus the full stability group of $\VEV{\chi}$ is $H_0 \oplus H_0'$ 
consisting of two disconnected continuous subsets,
where $H_0'$ are all the elements of the form $h (i J_{67})$ with $h\in H_0$. 
It may be noted that  $i J_{67}$ also enters 
the $D$-parity defined as $D\equiv (i J_{67})(i J_{23})$, the effective charge conjugation operator due to the charges of the fermions
assigned to $\mathbf{16}$.

In the model of \cite{Kibble:1982dd} the sequence of breaking is
\begin{align}
& Spin(10) {{M_X \atop{\longrightarrow}} \atop {\bf54}} Spin(6)\otimes Spin(4) 
{{M_R \atop{\longrightarrow}} \atop{{\bf126}}} 
SU(3)\otimes SU(2)\otimes U(1)\nonumber\\ & {{M_W \atop \longrightarrow} \atop {\bf10}} SU(2)\otimes U(1).
\end{align}
Stable cosmic strings arise at the first phase transition, due to $\Pi_0(H_0\oplus 
H'_0)= \mathbb{Z}_2$. At the next stage of breaking when the ${\bf126}$ acquires a VEV, domain walls 
appear due to the breaking of this $Z_2$. But the ${Z}_2$ comes embedded in
a continuous loop $U(1)_D$ the one-parameter subgroup generated by the $D$-parity generator. 
Such loops continuously connect a VEV to its $D$-parity conjugate.  Specifically such 
walls separate vacua with $\VEV{{\bf126}}=
(\bar{10},1,3)$ ( written in its components with Pati-Salam quantum numbers) from its 
charge conjugate VEV $(10,3,1)$.   
Thus the walls are not stable, and decay due to tension of the string boundaries which are 
liable
to shrink. The walls can also disintegrate due to creation of holes formed in them due to quantum 
tunnelling assisted by 
thermal fluctuations. If the second phase transition is first order, there is a phase of wall 
domination and the possibility of wall decay only through large black hole formation. This is 
certainly ruled out by the CMB inputs into primordial fluctuations. On the other hand, a second 
order phase transition at second  stage of breaking creates a short period of wall persistence 
though the walls do not come to dominate over radiation. 

These considerations are a warm up for the study of topology of the vaccum manifold 
arising in \cite{Aulakh:2003kg,Bajc:2004xe,Bajc:2004fj,Aulakh:2008sn} which study the breaking
of SUSY $SO(10)$ to MSSM. 
We argue 
that $D$-parity TPD walls are a necessary consequence in such a breaking and expect that an epoch 
similar to the second order phase transition at second stage as reviewed in this section may unfold.
A short period of substantial wall presence can have definite consequences to CMB data. We shall 
discuss this in Sec.s \ref{sec:DWso10} and \ref{sec:DWinfl}.

\section{D-parity symmetric vacua in left-right symmetric model}
\label{sec:LRDparity}
Before proceeding to the SUSY $SO(10)$ case, to 
illustrate the procedure we start with a warm up exercise for a related system, the minimal 
supersymmetric left-right symmetric model (MSLRM) considered in \cite{Aulakh:1997ba},
 based on the group $G_{LR} =$ $SU(3)_C \times 
SU(2)_L \times SU(2)_R\times U(1)_{B-L}$. While unlikely to have implications for inflation, the 
model is interesting in its own right as an intermediate scale group. The walls were studied from 
the point of view of cosmology and leptogenesis earlier in \cite{Mishra:2009mk, Mishra:2008be}.

The Higgs superfields proposed for breaking the $G_{LR}$ symmetry to SM are
\begin{align}
&\Delta=(1,3,1,2); \quad  \bar{\Delta}=(1,3,1,-2);\nonumber\\&
\Delta_c=(1,1,3,-2); \quad  \bar{\Delta}_c=(1,1,3,2); \nonumber\\&
\Omega=(1,3,1,0);\quad \Omega_c=(1,1,3,0). \label{fields}
\end{align}
These fields transform under D-parity as
\be 
\Delta \rightarrow  \Delta_c^*;\,\,\,\,\,\, \bar{\Delta} \rightarrow 
\bar{\Delta}_c^*;\,\,\,\,\,\,\, \Omega \rightarrow \Omega_c^*. 
\ee
The renormalizable superpotential corresponding to these Higgs superfields  is given as,
\bea 
W_{LR}&=&m_{\Delta}(Tr \Delta \bar{\Delta}+Tr \Delta_c \bar{\Delta_c})+m_{\Omega}(Tr \Omega^2+ 
Tr \Omega_c^2)\nonumber\\&&+a(Tr \Delta \Omega \bar{\Delta}+Tr \Delta_c \Omega_c \bar{\Delta_c}).
\eea
The vacua are sought assuming the supersymmetry to be unbroken and remaining so till the electroweak 
scale ($\sim$ O(TeV)). These can be obtained by imposing F-flatness and D-flatness 
conditions  given in \cite{Aulakh:1997ba}.
The set of vacuum expectation values (VEV's) for the Higgs fields required to obtain the MSSM is,
 \bea 
 \VEV{\Omega_c} &=& \begin{pmatrix}
  w_c & 0  \\
0 & -w_c   
\end{pmatrix}; \quad  \VEV{\Delta_c}= \begin{pmatrix}
  0 & 0  \\
d_c & 0   
\end{pmatrix}; \quad \VEV{\bar\Delta_c}= \begin{pmatrix}
  0 & \bar d_c  \\
0 & 0  
\end{pmatrix}; \nonumber \\&& \VEV{\Omega}=0; \quad  \VEV{\Delta}=0; \quad
\VEV{\bar\Delta}= 0.
\label{eq:leftlike}
\eea
The required minimum is obtained at $w=\frac{-m_{\Delta}}{a}$  and $d=(\frac{2 m_{\Delta} 
m_{\Omega}}{a^2})^{\frac{1}{2}}$  \cite{Aulakh:1997ba}. 
Here, $w$ and $d$ set two mass scales in the problem. 
At  first step, $\Omega_c$ acquires VEV at 
scale $M_R$ and $SU(2)_R$ is broken to $U(1)_R$,
and then  the 
VEV's of the $\Delta_c$, $\bar\Delta_c$ break  $U(1)_R \times U(1)_{B-L}$ to $U(1)_Y$ at a lower scale 
$M_{B-L}$. Thus, at this scale we get the minimal
supersymmetric standard model (MSSM).
However the $D$ and $F$-flatness conditions \cite{Aulakh:1997ba} also give another set of 
possibility of 
vacuum  which is degenerate to the one given by Eq. (\ref{eq:leftlike}) which
preserves the $SU(2)_R \times U(1)_L \times U(1)_{B-L}$ symmetry. The alternative set of VEV's is 
given by
\bea 
\VEV{\Omega} &=& \begin{pmatrix}
  w & 0  \\
0 & -w   
\end{pmatrix}; \quad  \VEV{\Delta}= \begin{pmatrix}
  0 & 0  \\
d & 0   
\end{pmatrix};  \quad \VEV{\bar\Delta}= \begin{pmatrix}
  0 & \bar d  \\
0 & 0  
\end{pmatrix}; \nonumber \\&& \VEV{\Omega_c}=0, \quad  \VEV{\Delta_c}=0, \quad
\VEV{\bar\Delta_c}= 0.
\label{eq:rightlike}
\eea
Due to the left-right symmetry, numerically $d=\bar d$, $d_c=\bar d_c$ and $w=w_c$. It is the breaking 
of this symmetry that leads to the formations of domain walls. Now we have 
two degenerate vacua separated by a domain wall. Since in this case the $D$-parity is a discrete 
symmetry,  the walls are topologically stable.
Here we consider an ansatz for a trajectory in the group space which connects the two 
vacua.  We parameterize the VEV's as follows with a parameter $\theta$,
  \begin{align} &\VEV{\Omega_c} = \cos \frac{\theta}{2}\begin{pmatrix}
  w_c & 0  \\
0 & -w_c   
\end{pmatrix}; \quad   \VEV{\Delta_c}= \cos \frac{\theta}{2}\begin{pmatrix}
  0 & 0  \\
d_c & 0   
\end{pmatrix};  \nonumber\\&  \VEV{\bar\Delta_c}=\cos \frac{\theta}{2} \begin{pmatrix}
  0 & \bar d_c  \\
0 & 0  
\end{pmatrix}; \quad
  \VEV{\Omega} = \sin \frac{\theta}{2}\begin{pmatrix}
  w & 0  \\
0 & -w   
\end{pmatrix}; \nonumber\\&   \VEV{\Delta}= \sin \frac{\theta}{2}\begin{pmatrix}
  0 & 0  \\
d & 0   
\end{pmatrix}; \quad \VEV{\bar\Delta}=\sin \frac{\theta}{2} \begin{pmatrix}
  0 & \bar d  \\
0 & 0
\end{pmatrix}.\end{align}
When $\theta$=0, we have left like vacuum and for $\theta$ = $\pi$, right like. On substituting 
these paramterised VEV's in the superpotential we obtain,
 \bea 
 W_L= &m_{\Delta} \cos^2\frac{\theta}{2} d_c^2+2 m_{\Omega} \cos^2\frac{\theta}{2} w_c^2+a 
\cos^3\frac{\theta}{2} d_c^2 w_c\nonumber\\
 W_R= &m_{\Delta} \sin^2\frac{\theta}{2} d^2+2 m_{\Omega} \sin^2\frac{\theta}{2} w^2+a 
\sin^3\frac{\theta}{2} d^2 w. 
\eea
We can then compute $\theta$ derivative of the scalar potential as
 \be
 \frac{\partial V}{\partial \theta}= 2 Re \sum_{i} \frac{\delta W}{\delta \phi_i} 
\frac{\partial}{\partial \theta} (\frac{\delta W}{\delta \phi_i}).
  \ee
  This gives, using the numerical equality of the VEV's noted below Eq. (\ref{eq:rightlike}),
\bea
\frac{\delta V_{total}}{\delta \theta}&=& -\sin\theta[\cos^2 \frac{\theta}{2} \{(m_{\Delta} d_c+ 
a \cos\frac{\theta}{2} d_c w_c) \nonumber\\&&(m_{\Delta} d_c+ \frac{3a}{2} \cos\frac{\theta}{2} d_c 
w_c)+ (2 m_{\Omega} w_c+a \cos\frac{\theta}{2} d_c^2)\nonumber\\&& (2 m_{\Omega} w_c+  \frac{3a}{2} 
\cos\frac{\theta}{2} d_c^2)\}
- \sin^2 \frac{\theta}{2} \{(m_{\Delta} d_c+ a \sin\frac{\theta}{2} d_c w_c) \nonumber\\&&(m_{\Delta} d_c+ 
\frac{3a}{2} \sin\frac{\theta}{2} d_c w_c)+ (2 m_{\Omega} w_c+a \sin\frac{\theta}{2} 
d_c^2) \nonumber\\&&(2 m_{\Omega} w_c+  \frac{3a}{2} \sin\frac{\theta}{2} d_c^2) \}].
\eea
It is easy to see that the two expressions with the braces mutually cancel
at the symmetric point $\theta$=$\frac{\pi}{2}$. The value of 
the energy at this point is given by,
\bea V_{DW}= (2-\sqrt{2})^2 \frac{m_{\Delta} m_{\Omega}}{a^2} (m_{\Delta}^2+m_{\Omega}^2). \eea 

The two set of vacua considered above, Eqs. (\ref{eq:leftlike}), (\ref{eq:rightlike}) are degenerate 
and related by $D$-parity, which is a discrete symmetry of the group $G_{LR}$. The domain walls are therefore topologically 
stable. The solitonic domain walls thus arising were obtained as solutions of this theory in \cite{Sarkar:2007er}. 
The motivation here is different.  The considerations of this section illustrate how one rotates from one vacuum to 
another, not necessarily along energetically optimal path, but in order to estimate the height of the barrier. 
The same strategy will be utilised even for the more general case when the parent group is simply connected.
The main point is that degeneracies that might occur in single field minimization are lifted due to the presence of several mutually coupled fields 
providing a general quartic polynomial.
The two minima of Eqs. (\ref{eq:leftlike}) and (\ref{eq:rightlike}) are two of the solutions of the extremization condition. 
Such extrema are necessarily isolated points, being the zeros of a generic polynomial. Further, supersymmetry   ensures that 
the supersymmetry preserving  minima are absolute minima. Thus the third extremum, the intermediate point, is a local maximum determined along the 
parameterized curve. While this is not guaranteed to be the lowest energy peak separating the two minima, it provides 
an upper bound on the height of the saddle point lying on the barrier.

\section{Domain walls in minimal supersymmetric SO(10) GUT}
\label{sec:DWso10}
We now turn to the main problem of the minimal SUSY GUT model (MSGUT)  
\cite{Aulakh:2003kg,Aulakh:1982sw,Babu:1992ia,Bajc:2004fj}. 
The wall ansatz for MSGUT turns out to 
be complicated due to the presence of several Higgs fields 
in different representations. At first we proceed to establish the presence of a variety 
of walls that may appear in this MSGUT model, which may also involve intermediate phases of 
smaller gauge groups,  if we allow some variation in the parameters. In each case we focus on 
$D$-parity walls which are unstable yet have non-trivial consequences.

The MSGUT can be broken down to MSSM directly or through intermediate symmetries depending 
upon the choice of Higgs multiplet getting VEV \cite{Bajc:2004xe}.  During these symmetry breakings, 
the $D$-parity is also broken and leads to the formation of TPD domain walls.
The Higgs content of this model is ${\bf{210}}$ ($\Phi_{ijlk}$, 
four index totally antisymmetric), $\bf{126}(\bf{\overline{126}})$ ($\Sigma_{ijklm}$ 
($\overline\Sigma_{ijklm}$),  five index totally 
anti-symmetric self-dual (anti-self-dual) representation) and the vector representation {\bf 10} ($H_i$).  Here 
$i,j,k,l,m =1,2...10 $ run over the vector representation of $SO(10)$. 
The \textbf{126}($\overline{\textbf{126}}$) and  $\textbf{210}$
 break the SO(10) gauge symmetry to MSSM; the $\mathbf{10}$ breaks the electroweak symmetry,
 while the $\mathbf{10}$ and $\overline{\textbf{126}}$ 
give masses to the fermions.

The  renormalizable superpotential for the above mentioned Higgs superfields is given by,
  \bea 
  W&=&\frac{m_{\Phi}}{4!} \Phi^2+\frac{\lambda}{4!} \Phi^3 + \frac{m_{\Sigma}}{5!}
  \Sigma \overline{\Sigma}+\frac{\eta}{4!} \Phi \Sigma \overline\Sigma+m_{H} H^2\nonumber\\&&
  +\frac{1}{4!} \Phi H (\gamma \Sigma+\bar \gamma \overline\Sigma)\,.
  \eea
To recognize the  SM singlets, the decomposition of Higgs supermultiplets required for SO(10) 
symmetry breaking to 
  MSSM in terms of Pati-Salam gauge group
  ($SU(4)_C \times SU(2)_L \times SU(2)_R$) is given as  \cite{Bajc:2004xe},
\bea
210 &= (15,1,1)+(1,1,1)+(15,1,3)+(15,3,1)\nonumber\\&+(6,2,2)+(10,2,2)+(\bar{10},2,2)\nonumber\\
  126 &=(\bar{10},1,3)+(10,3,1)+(6,1,1)+(15,2,2)\nonumber\\
 \overline{126} &=(\bar{10},3,1)+(10,1,3)+(6,1,1)+(15,2,2)\nonumber
  \eea
So, we call the SM singlet fields as $P$(1,1,0), $A$(irreducible singlet of (15,1,1))
and $\Omega_R^{0}$( $(113^0)$ of (15,1,3)) from ${\bf 210}$. Similarly  we identify $\Sigma_{R}^{-}$ ($({\bar1}13^-)$ 
of $(\bar 10, 1, 3)$) from ${\bf 126}$ and $\overline\Sigma_{R}^{+}$ ( $({\bar1}13^+)$ 
of $( 10, 1, 3)$) from $\overline{\bf 126}$.
The details of how these fields are defined in terms of components having SO(10) indices breaking them in $SO(6)\otimes SO(4)$ indices is elaborated in 
appendix \ref{sec:appA}. The VEV of $H$ is not relevant to our considerations.
The D-parity is defined as
\begin{equation}
D= exp(i \pi J_{23})exp(i \pi J_{67})
\end{equation}
Under the action of D-parity these fields transform as
\bea
P \rightarrow -P; \,\,\,\,\,\,
A \rightarrow A;\,\,\,\,\,\,\,
W_R^{0} \rightarrow W^{0}_{L} \nonumber\\
\Sigma_R^{-} \rightarrow -\Sigma^{+}_{L} ;\,\,\,\,\,\,\,\,\,
\overline{\Sigma}_R^{+} \rightarrow -\overline{\Sigma}^{-}_{L}
\eea 
as further explained in the appendix \ref{sec:appA}.
Specific components of these fields are assigned the following VEV's.
\begin{align} 
&\VEV{\Phi_{78910} }= p \nonumber\\&
 \VEV{\Phi_{1234}}= \VEV{\Phi_{1256}}= \VEV{\Phi_{3456}}=a\nonumber\\&
 \VEV{ \Phi_{1278}}=\VEV{ \Phi_{3478}}=\VEV{ \Phi_{5678}}\nonumber\\&
 =\VEV{ \Phi_{12910}}=\VEV{ \Phi_{34910}}=\VEV{ \Phi_{56910}}=w\nonumber\\&
 \VEV{\Sigma_{a+1,b+3,c+5,d+7,e+9}}=\frac{1}{2^{5/2}} (i)^{a+b+c-d-e} \sigma \nonumber\\&
 \VEV{\overline \Sigma_{a+1,b+3,c+5,d+7,e+9}}=\frac{1}{2^{5/2}} (-i)^{a+b+c-d-e} \bar\sigma
 \label{eq:vevchoice}
\end{align}
 so that, $\VEV{\Omega^{0}_L}$ = $\VEV{\overline{\Sigma}^{-}_{L}}$=$\VEV{\Sigma^{+}_{L}}$=0. 
 Here $a,b,c,d,e$ take values $0$ or $1$. 

The superpotential in terms of these VEVs is given by
\bea
W&=&m_{\Phi} (p^2+a^2+w^2)+
2\lambda (a^3+3pw^2+6aw^2) \nonumber\\&&
 +m_{\Sigma}\sigma \overline{\sigma}
 +\eta \sigma \overline{\sigma}(p+3a+6w).\eea
 The SUSY preserving minima  
 using the $F$-term and $D$-terms vanishing conditions are given by \cite{Aulakh:2003kg},   
 \bea
a= \frac{m_{\Phi}}{\lambda}\frac{x^2+2x-1}{1-x};\, p=\frac{m_{\Phi}}{\lambda} \frac{x(5
x^2-1)}{(1-x)^2};\,\nonumber\\
\sigma\overline{\sigma}=\frac{2 m_{\Phi}^2}{\eta \lambda} \frac{x (1-3 x)(1+x^2)}{\eta (1-x)^2};\,\, 
w = -\frac{m_\Phi}{\lambda} x. 
\label{eq:mssm}
\eea
where $x$ is the solution of following cubic equation
 \bea 
 8 x^3-15 x^2+14 x -3 = -\frac{\lambda m_{\Sigma}}{\eta m_\Phi} (1-x)^2 .
 \label{eq:xcubic}
 \eea
 However we have a list of possible intermediate symmetries depending on the value of $x$ \cite{Bajc:2004xe}. 
  \begin{enumerate}
 \item For $x=1/2$ and if $\lambda m_{\Sigma}/\eta m_{\Phi}$ = $-5$, it gives SU(5) minimum.
  \item For $x=0$ and if $\lambda m_{\Sigma}/\eta m_{\Phi}$ = $3$, this results in $SU(3)_C \times 
SU(2)_L \times SU(2)_R \times U(1)_{B-L}$ minimum.
 \item For $x=\pm i$ and if $\lambda m_{\Sigma}/\eta m_{\Phi}$ = $-3$ ($1 \pm 2i$), it gives $ SU(3)_C 
\times SU(2)_L \times U(1)_R \times U(1)_{B-L}$ symmetry.
  \item For $x=1/3$ and if $\lambda m_{\Sigma}/\eta m_{\Phi}$ = $-2/3$, it results in the flipped $SU(5) 
\times U(1)$ minimum.
  \item For $x=1/4$ and if $\lambda m_{\Sigma}/\eta m_{\Phi}$ = $5/9$, it results in MSSM minimum.
   \end{enumerate}
   Now, consider an arbitrary $D$-rotation  
   \bea 
U(\theta)_D= exp\lbrace i \theta (J_{23}+J_{67})\rbrace.
\eea
Individual components of the fields  transform differently under this generalized $U_D$-rotation, as follows, (with 
$s_{\theta}$, $c_{\theta}$ standing for $\sin\theta$ and $\cos\theta$ respectively)
 \begin{align}
&\hat{\Phi}_{78910}=c_{\theta}\Phi_{78910}+ s_{\theta}\Phi_{68910}  \nonumber\\&
\hat{\Phi}_{1234} = \Phi_{1234}\nonumber\\&
\hat{\Phi}_{1256}= c_{\theta}^2 \Phi_{1256}- c_{\theta}s_{\theta}(\Phi_{1356}+ 
\Phi_{1257})+s_{\theta}^2\Phi_{1357} \nonumber\\&
\hat{\Phi}_{3456}= c_{\theta}^2 \Phi_{3456}+ c_{\theta}s_{\theta}(\Phi_{2456}- 
\Phi_{3457})-s_{\theta}^2\Phi_{2457} \nonumber\\&
\hat{\Phi}_{1278}= c_{\theta}^2 \Phi_{1278}+ c_{\theta}s_{\theta}(\Phi_{1378}- 
\Phi_{1268})-s_{\theta}^2\Phi_{1368} \nonumber\\&
\hat{\Phi}_{3478}= c_{\theta}^2 \Phi_{3478}+ c_{\theta}s_{\theta}(\Phi_{2478}+ 
\Phi_{3468})+s_{\theta}^2\Phi_{2468} \nonumber\\&
\hat{\Phi}_{5678}= W_{5678} \nonumber\\&
\hat{\Phi}_{12910}=c_{\theta}\Phi_{12910}-s_{\theta}\Phi_{13910}  \nonumber\\&
\hat{\Phi}_{34910}=c_{\theta}\Phi_{34910}+ s_{\theta}\Phi_{24910}  \nonumber\\&
\hat{\Phi}_{56910}=c_{\theta}\Phi_{56910}- s_{\theta}\Phi_{57910} \nonumber\\&
\hat{\Sigma}_{13579}= c_{\theta}^2 \Sigma_{13579}+ c_{\theta}s_{\theta}(\Sigma_{12579}+ 
\Sigma_{13569})+s_{\theta}^2\Sigma_{12569} \nonumber\\
\end{align}
Similarly one can write out for the other field components of $\Sigma^{-}_R$ and 
$\overline{\Sigma}^+_R$  given in \ref{LComponents}. 

\begin{figure}[t]
\begin{subfigure}{.5\textwidth}
  \includegraphics[width=.8\linewidth]{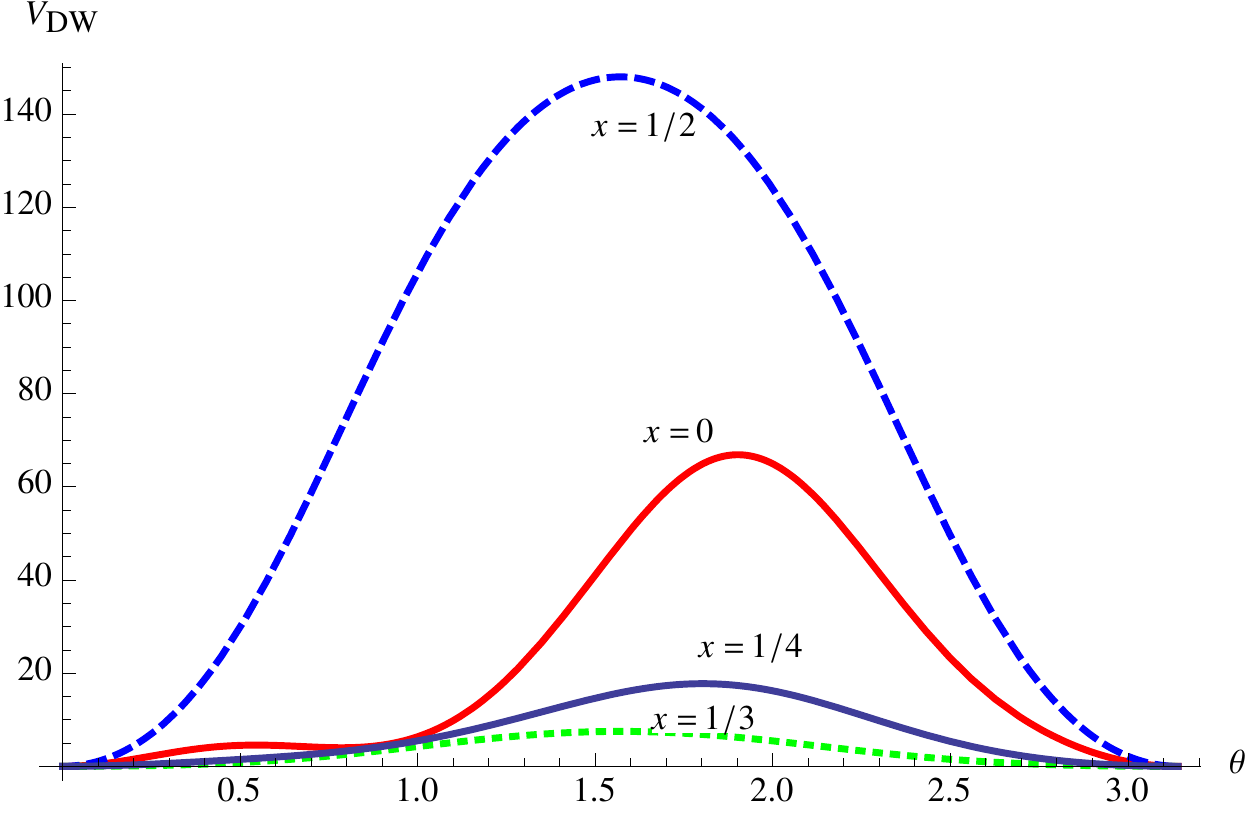}
  \caption{}
  \label{fig:sub1}
\end{subfigure}
\begin{subfigure}{.5\textwidth}
  \includegraphics[width=.8\linewidth]{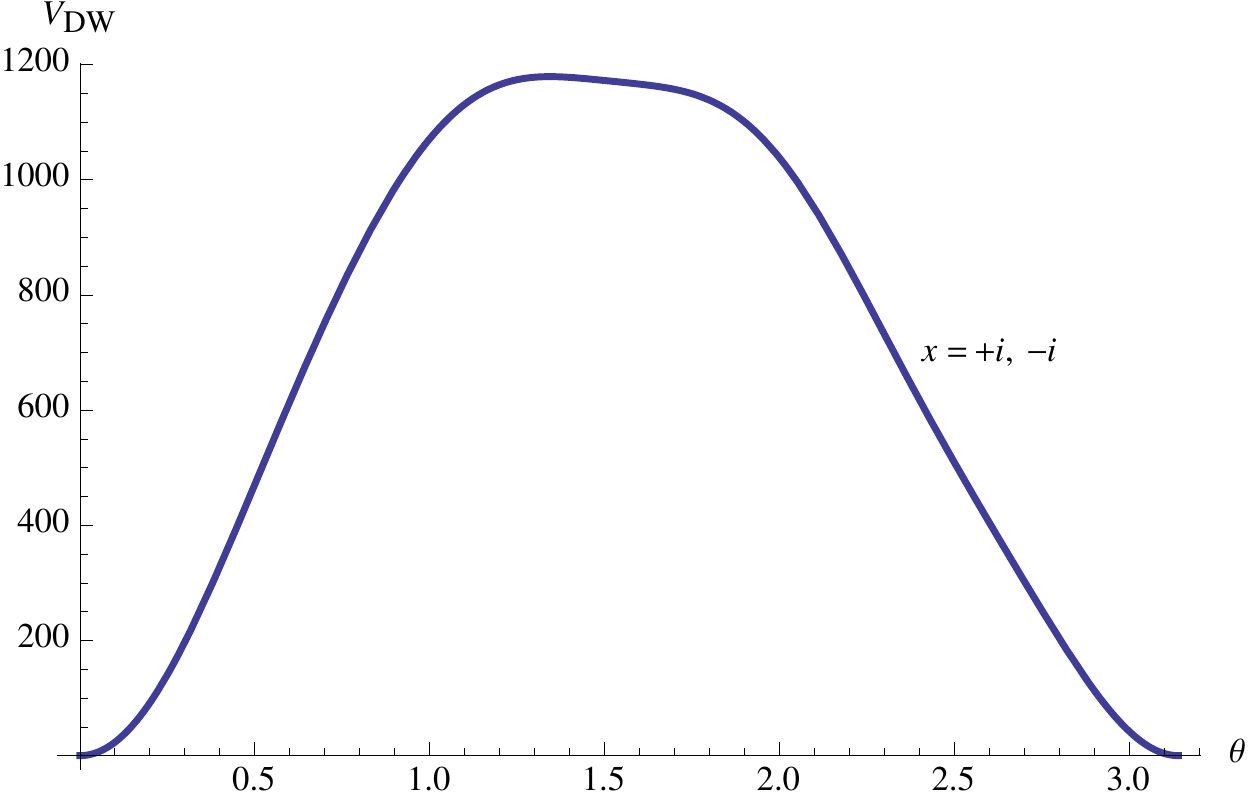}
  \caption{}
  \label{fig:sub2}
\end{subfigure}
\caption{ The two subfigures (a) and (b) show scalar potential in arbitrary units as a function of $\theta$ which generates a one parameter subgroup. 
The curves correspond to  different patters of symmetry breaking labeled by $x$, as listed below Eq. (\ref{eq:xcubic}). The one step breaking to MSSM corresponds to 
  $x=1/4$.}
  \label{potvstheta}
\end{figure}

Now we calculate the $\theta$ dependent potential from the corresponding superpotential as,
\bea V= \sum_{i=1}^{74} \biggr| \frac{\partial W}{\partial \phi_i}\biggr|^2 
\label{eq:PF}\,.
\eea
Here $i$  runs over number of field components given in Eq. (\ref{LComponents}). 
The form of potential for different values of $x$ assuming $|\eta|=|\lambda|$ is 
\begin{align}
& V^{DW}_{x=0}=\frac{| m_{\Phi}| ^4}{| \lambda| ^2} (8 ( \cos 2 \theta +\sin 2 \theta -1) ^2
\nonumber\\&+( -2 \cos 2 \theta +\sin 4 \theta +2)
   ^2)\nonumber\\&
V^{DW}_{x=1/3}=   \frac{16 | m_{\Phi }| ^4 (26 \sin^4 \theta+12 \sin ^2\theta)}{81 | \lambda| ^2}\nonumber\\&
V^{DW}_{x=\pm i}=\frac{|m_{\Phi}| ^4 }{| \lambda |^2}(272 \sin ^4\theta+160 \sin ^2\theta \nonumber\\&
+48 \sin ^2\theta (4 \sin 2 \theta +11 \cos 2 \theta +25))\nonumber\\&
V^{DW}_{x=1/2}=  \frac{|m_\Phi| ^4\sin ^2\theta}{8 |\lambda| ^2}  (-159 \cos 2 \theta -5 (14 \cos 4 \theta \nonumber\\& +\cos 6 \theta -218))\nonumber\\&
V^{DW}_{x=1/4}=   \frac{|m_\Phi |^4 \sin ^2 \theta}{93312 | \lambda| ^2} (-490680 \sin 2 \theta +111780 \sin 4 \theta \nonumber\\& -324597 \cos 2 \theta+41142 \cos 4 \theta +17613 \cos 6  \theta +1127498)
\end{align}
The variation of the potential (in units of $\frac{|m_\Phi| ^4}{ |\lambda| ^2}$) as a function of the $U(1)_D$-rotation angle $\theta$ is shown in Fig. \ref{potvstheta}. One can 
see from Fig.  \ref{potvstheta}  
that at $\theta$ = 0 and $\pi$, the potential energy is zero and 
the VEVs satisfy the relations in Eq. (\ref{eq:mssm}) which are $x$ dependent. 
For intermediate values of $\theta$ the potential is not symmetric under $\theta \rightarrow \pi-\theta $ and its overall magnitude is also
strongly $x$ dependent.

Our main motivation in performing this detailed calculation is to establish that TPD domain walls 
indeed form even if the group contains no discrete symmetry. In the last para of Sec. 
\ref{sec:LRDparity}, it was pointed out that our strategy at least yields an upper bound on the 
energy barrier separating vacua which are two distinct points. Unlike in that example, the group is 
simply connected here. However, it may be observed that 
there is an inadvertent (not accidental) discrete symmetry of the $D$ and $F$ 
flatness conditions. The flatness conditions cannot single out a unique vacuum 
but signal two for a given set of parameters, related by the $D$-parity. We now need to argue that 
this vacuum pair related by the flip symmetry are 
necessarily separated by an energy barrier. It is sufficient to focus on the $210$ whose three independent
sets of MSSM singlet components are assigned different VEV's $p$, $a$ and $w$. The parameters in the 
superpotential are tuned according to Eq.s \eqref{eq:mssm} \eqref{eq:xcubic} and the five possibilities listed 
below them. Then the value $x=1/4$ is one zero of a cubic polynomial, which is necessarily isolated.  Further,
any values of the component fields $P$, $A$ and $\Omega$ accessed by small variations of $p$, $a$ and $w$ 
are necessarily of higher energy.  Thus the preferred vacua with unbroken MSSM are also isolated points at 
best connected by discrete transformations.  
This ends the existence proof of isolated vacua.  While 
it is convenient to 
build low energy phenomenology based on the preferred vacuum, the conditions in the early universe 
allow domains of both types to form. 
Eventually the unstable TPD walls must disintegrate or have 
unfavorable consequences as discussed in \ref{sec:KLSreview}, based on \cite{Kibble:1982dd}.
In the next section we turn to cosmological consequences for MSGUT.

\section{Topological pseudo-defect walls and inflation}      
\label{sec:DWinfl}
It is interesting to inquire what kind of signatures the walls can leave. A high scale
theory will necessarily have to contend with inflation scale physics. Broadly, we may
consider three possibilities, (A) $M_{GUT} < M_{inf}$, (B) $M_{GUT} > M_{inf}$, (C) $M_{GUT} \simeq
M_{inf}$, where $\mgut$ is the $SO(10)$ symmetry breaking scale and $M_{inf}$ is the scale of 
inflation.

Case A is generic to chaotic inflation\cite{Linde:1983gd} where inflation originates close to the 
Planck scale. In this case after reheating, the temperature 
could be less than or more than $\mgut$. In the former
case the thermal state should be  directly in the required MSSM phase. In the
latter case however, after the symmetry breaking phase transition, TPD 
walls would emerge. Due to their unstable
nature they eventually disintegrate. 
The resulting epoch of wall domination
would end with  entropy dumping with return to pure radiation dominated
universe. It has been pointed out \cite{Martin:2010kz}\cite{Dai:2014jja}\cite{Cook:2015vqa} that there 
are models of inflation in which the duration of the reheating phase and the effective equation of state 
during that phase can be correlated with other inflation observables and is being pursued in \cite{Rajesh:2018}.
For such cases the presence of TPD walls during reheating could have important consequences.

In Case B, the TPD walls would be copiously present when inflation commences. This would produce 
signatures similar in nature to but more pronounced than in the case C to be discussed below. The
epoch over which the power law inflation caused by the walls would compete with the scale 
invariant inflation would be determined by the ratio $\mgut/M_{inf}$. In late time observables, 
this would reflect in deviations from scale invariance at the largest scales. Since there are no
strong indications to this effect we do not analyse this further, however the framework would be 
similar to that we pursue for the case C.

Case C could be accidental, but more interestingly, also occurs if inflationary physics emerges
from the same Grand unified theory. In this
case the formation of TPD walls could occur before the inflaton potential energy
 dominates, giving rise to the signatures  in the primordial fluctuations 
as encoded in the CMB data\cite{Smoot:1992td,Bennett:2012zja}. Recently this question has been 
addressed in 
\cite{BouhmadiLopez:2012by} and it is shown that the presence of frustrated domain walls can 
alleviate the quadrupole anomaly of the CMB fit occurring in the Lambda-CDM model.
 
Consider the presence of domain walls which are conformally stretched, 
\cite{Kibble:1982dd} and the wall complex as a whole obeys the coarse grained equation of state
\cite{Kolb:1990vq}, and corresponding dependence on the Friedmann scale factor,
\be
p=-\frac{2}{3}\rho; \qquad \rho_{\scriptstyle{DW}}(t)=\frac{\rho_1 a_1}{a(t)}.
\ee
where $\rho_1\equiv \mgut^4$, and in the latter equation the numerator sets the initial conditions 
on its value.
The inflaton has a comparable energy density, $V_0\equiv M_{inf}^4$ so that $H_0^2=(8\pi/3)G V_0$ 
would be the Hubble parameter if only the inflaton were present. The 
combined Friedmann 
equation,
\be
\left(\frac{\dot{a}}{a}\right)^2 = \frac{8\pi}{3}G \left(V_0 + \frac{\rho_1 a_1}{a(t)}  \right),
\label{eq:DWfriedmann}
\ee
has the solution
\be
a(t) = \frac{\rho_1 a_1}{2 V_0} \left[\cosh \left\lbrace H_0(t-t_1)+u_1 \right\rbrace -1  \right], 
\ee
with
\be
\cosh u_1=1+\frac{2V_0}{\rho_1}.
\ee
In the regime where $\rho_1 > V_0$, and for $H_0(t-t_1)<1 $ one gets the behaviour
\be
a^{(1)}(t) \approx \frac{4\pi}{3}G \rho_1 a_1 \left( 1+\frac{2V_0}{\rho_1} \right)(t-t_1)^2,
\ee
characteristic of the $p=-2\rho/3$ equation of state. At late times of course the vacuum energy 
dominates. But a brief period of wall domination would 
still have the behavior similar to inflation, in which physical scales like those of the
scalar field perturbations would be growing faster than the Hubble horizon. The amplitude of the 
perturbations would be imprinted on 
the earliest of the scales to leave the horizon.  The $\epsilon$ parameter of inflation calculated
in the present case gives,
\be
\epsilon=-\frac{\dot{H}}{H^2}=\frac{1}{2}\textrm{sech}^2[\frac{1}{2}\lbrace H_0(t-t_1)+u_1 \rbrace]. 
\ee
At $t=t_1$ this gives $\epsilon=1/2$ as expected for a pure power law expansion with domain walls. 
But it soon turns over to 
\be
\epsilon \approx \frac{1}{2}e^{-H_0(t-t_1)},
\ee
approaching the value $0$ of the vacuum energy dominated phase. 
Thus the early modes to leave the horizon would be far from scale invariant, whereas within a few 
e-foldings of the time scale $H_0^{-1}$ the slow roll condition is satisfied \cite{Rajesh:2018}. 
The departure from approximate scale invariance could therefore be detected. Further, the presence 
of domain walls would introduce non-gaussianities. Since these would affect inflation only in its 
earliest stages of slow roll, they may not have entered our horizon yet. But in principle these 
could be detected. While cosmic strings have been studied for their effect on CMB data extensively
\cite{Durrer:2001cg}\cite{Urrestilla:2007sf}\cite{Ringeval:2010ca}\cite{Ade:2013xla}, 
the presence of such primordial domain walls is also warranted, as a countercheck on models of 
unification as well as inflation.

\section{Conclusions}
\label{sec:concl}
We have studied the example of a unification group wherein domain walls can form 
although the group is simply connected, with no discrete symmetries that break spontaneously. But 
inadvertent symmetries 
of the minimization conditions imply the possibility of a discrete set of vacua. Such 
vacua turn out to be related by discrete symmetries in the parent group. 
The case in point is the well known $D$-parity of $Spin(10)$ and its 
descendants. We have explicitly computed value of the energy along one-parameter paths connecting 
two possible subgroups to which the symmetry breaking of $Spin(10)$ could have occurred. We 
thus show that the vacua are indeed separated by an energy barrier. Then 
the causal structure of the early Universe creates the interesting possibility of
topological pseudodefects, dubbed TPD walls, separating regions of such vacua. Even though 
manifestly unstable, the walls may 
live long enough to leave  imprints on the observables. Such signatures in the CMB signals 
create the exciting possibility of accessing grand unification in current observations. 

We have shown that in the context of inflation with a preceding epoch of radiation 
domination, (cases B and C), the formation of domain walls would leave scale dependent imprints on 
the very long 
wavelengths which leave the horizon at the onset of inflation. At current state of knowledge we do 
not know 
if these are indeed the scales being seen in the lowest multipoles. Likewise it is important to study 
non-gaussianities resulting from such objects in the phase at the onset of inflation.

\appendix
\section{}
\label{sec:appA}
The MSSM singlets components  from ${\bf210}$ are $(15,1,1)$, $(1,1,1)$, $(15,1,3)$, each of which is
assigned a different VEV. Further, we have $(\bar{10},1,3)$ from ${\bf126}$ and  
$(10,3,1)$  from $\overline{\bf 126}$, all of which acquire VEVs to break $SO(10)$ down to SM. These can 
be written in terms of $SO(10)$ vector indices. We follow the procedure given in \cite{Bajc:2004xe} 
but our conventions are different. We choose 
$a,b=1,2,...6$ for $SO(6)$ and $\alpha$, $\beta$ $=7,8,9,10$ for $SO(4)$. Now the PS group $SU(4)_C 
\times 
SU(2)_L \times SU(2)_R$
is isomorphic to $SO(6)$ $\times$ $SO(4)$ $\subset$ $SO(10)$. In \cite{Bajc:2004xe}, the full table 
of 
Higgs representations 
in terms of $SO(10)$ indices is given. Below are the fields given in terms of $SO(10)$ indices in 
our 
conventions
which are important to us in terms of breaking of $SO(10)$ gauge group.
\bea
P&=&[7,8,9,10]\nonumber\\
A&=&[1,2,3,4]+[1,2,5,6]+[3,4,5,6]\nonumber\\
\Omega_R^{0}&=&[1,2,7,8]+[3,4,7,8]+[5,6,7,8]+[1,2,9,10] \nonumber\\&&+[3,4,9,10]+[5,6,9,10]\nonumber\\
\Sigma_{R}^{-}&=& -i([1,3,5,7,9]-[2,4,5,7,9]-[2,3,6,7,9]-[1,4,6,7,9]\nonumber\\&&-
i[2,3,5,7,9]-i[1,4,5,7,9]-i[1,3,6,7,9]+i[2,4,6,7,9])\nonumber\\&&-(7,9 \rightarrow 8,10)
+i\{7,9 \rightarrow 7,10\}+i\{7,9 \rightarrow 8,9\}\nonumber\\
\overline{\Sigma}_{R}^{+}&=& i([1,3,5,7,9]-[2,4,5,7,9]-[2,3,6,7,9]-[1,4,6,7,9]\nonumber\\&&
+i[2,3,5,7,9]+i[1,4,5,7,9]+i[1,3,6,7,9]-i[2,4,6,7,9])\nonumber\\&&-(7,9 \rightarrow 8,10)
-i\{7,9 \rightarrow 7,10\}-i\{7,9 \rightarrow 8,9\}
\label{LComponents}\eea
The sign $+(-)$ in the superscript represents the $T_{3R}$ value. 
The D-parity is defined as
\begin{equation}
D= exp(i \pi J_{23})exp(i \pi J_{67})
\end{equation}
Using the definition of MSSM singlet fields  in Eq. \ref{LComponents}, we find that under the action of D-parity these fields transform as
\bea
P \rightarrow -P; \,\,\,\,\,\,
A \rightarrow A;\,\,\,\,\,\,\,
W_R^{0} \rightarrow W^{0}_{L} \nonumber\\
\Sigma_R^{-} \rightarrow -\Sigma^{+}_{L} ;\,\,\,\,\,\,\,\,\,
\overline{\Sigma}_R^{+} \rightarrow -\overline{\Sigma}^{-}_{L}
\eea 
where,
\bea
\Omega^{0}_L&=&[7,8,1,2]+[7,8,3,4]+[7,8,5,6]-[9,10,1,2] \nonumber\\&&-[9,10,3,4]-[9,10,5,6]\nonumber\\
\overline{\Sigma}^{-}_{L}&=& -i([7,9,1,3,5]-[7,9,2,4,5]-[7,9,2,3,6]-[7,9,1,4,6]\nonumber\\&&-
i[7,9,2,3,5]-i[7,9,1,4,5]-i[7,9,1,3,6]+i[7,9,2,4,6])\nonumber\\&&+(7,9 \rightarrow 8,10)
-i\{7,9 \rightarrow 7,10\}+i\{7,9 \rightarrow 8,9\}\nonumber\\
\Sigma^{+}_{L}&=& i([7,9,1,3,5]-[7,9,2,4,5]-[7,9,2,3,6]-[7,9,1,4,6]\nonumber\\&&
+i[7,9,2,3,5]+i[7,9,1,4,5]+i[7,9,1,3,6]-i[7,9,2,4,6])\nonumber\\&&+(7,9 \rightarrow 8,10)
+i\{7,9 \rightarrow 7,10\}-i\{7,9 \rightarrow 8,9\}
\eea
The sign $+(-)$ in the superscript represents the $T_{3L}$ value. These are used to choose the VEV's used in Eq. (\ref{eq:vevchoice})

Next, the choice of the directions of the VEV's  for the D-rotated field components used 
in calculating the potential in Eq. (\ref{eq:PF}) 
is made as follows,
\bea &&\VEV{\Phi_{78910} } = \VEV{ \Phi_{68910} }=p \nonumber\\&&
 \VEV{\Phi_{1234}}= \VEV{\Phi_{1256}}= \VEV{\Phi_{3456}}= \VEV{\Phi_{1356}}= \VEV{\Phi_{1257}}= 
\VEV{\Phi_{1357}} 
 \nonumber\\&& = \VEV{\Phi_{2456}}= \VEV{\Phi_{3457}}= \VEV{\Phi_{2457}}=a\nonumber\\&&
 \VEV{ \Phi_{1278}}=\VEV{ \Phi_{3478}}=\VEV{ \Phi_{5678}}=\VEV{ \Phi_{12910}}=\VEV{ 
\Phi_{34910}}=\VEV{ \Phi_{56910}} \nonumber\\&& =\VEV{ \Phi_{1378}}=\VEV{ \Phi_{2478}}=\VEV{ 
\Phi_{13910}}=\VEV{ \Phi_{24910}}=\VEV{ \Phi_{57910}} \nonumber\\&&
=\VEV{ \Phi_{1268}}=\VEV{ \Phi_{1368}}=\VEV{ \Phi_{3468}}=\VEV{ \Phi_{2468}}=w \nonumber\\&&
\VEV{\Sigma_{13579}}=\VEV{\Sigma_{12579}}=\VEV{\Sigma_{13569}}=\VEV{\Sigma_{13569}}=\VEV{\Sigma_{
12569}}=\sigma \nonumber\\
\eea
Similarly we can write for all the field components which 
will appear after D-rotation of $\Sigma_{R}^{-}$ and $\overline{\Sigma}^+_R$ 
( taking care of the $i$ for each component of  $\Sigma_{R}^{-}$ and 
$\overline{\Sigma}^+_R$ appearing in Eq. \ref{eq:vevchoice}). \\

\bibliographystyle{apsrev}
\bibliography{references}

\begin{thebibliography}{48}
\expandafter\ifx\csname natexlab\endcsname\relax\def\natexlab#1{#1}\fi
\expandafter\ifx\csname bibnamefont\endcsname\relax
  \def\bibnamefont#1{#1}\fi
\expandafter\ifx\csname bibfnamefont\endcsname\relax
  \def\bibfnamefont#1{#1}\fi
\expandafter\ifx\csname citenamefont\endcsname\relax
  \def\citenamefont#1{#1}\fi
\expandafter\ifx\csname url\endcsname\relax
  \def\url#1{\texttt{#1}}\fi
\expandafter\ifx\csname urlprefix\endcsname\relax\def\urlprefix{URL }\fi
\providecommand{\bibinfo}[2]{#2}
\providecommand{\eprint}[2][]{\url{#2}}

\bibitem[{\citenamefont{Hirata et~al.}(1989)}]{Hirata:1989zj}
\bibinfo{author}{\bibfnamefont{K.~S.} \bibnamefont{Hirata}}
  \bibnamefont{et~al.} (\bibinfo{collaboration}{Kamiokande-II}),
  \bibinfo{journal}{Phys. Rev. Lett.} \textbf{\bibinfo{volume}{63}},
  \bibinfo{pages}{16} (\bibinfo{year}{1989}).

\bibitem[{\citenamefont{Fukuda et~al.}(2001)}]{Fukuda:2001nj}
\bibinfo{author}{\bibfnamefont{S.}~\bibnamefont{Fukuda}} \bibnamefont{et~al.}
  (\bibinfo{collaboration}{Super-Kamiokande}), \bibinfo{journal}{Phys. Rev.
  Lett.} \textbf{\bibinfo{volume}{86}}, \bibinfo{pages}{5651}
  (\bibinfo{year}{2001}), \eprint{hep-ex/0103032}.

\bibitem[{\citenamefont{Minkowski}(1977)}]{Minkowski:1977sc}
\bibinfo{author}{\bibfnamefont{P.}~\bibnamefont{Minkowski}},
  \bibinfo{journal}{Phys.Lett.} \textbf{\bibinfo{volume}{B67}},
  \bibinfo{pages}{421} (\bibinfo{year}{1977}).

\bibitem[{\citenamefont{Yanagida}(1979)}]{Yanagida:1979as}
\bibinfo{author}{\bibfnamefont{T.}~\bibnamefont{Yanagida}},
  \bibinfo{journal}{Conf. Proc.} \textbf{\bibinfo{volume}{C7902131}},
  \bibinfo{pages}{95} (\bibinfo{year}{1979}).

\bibitem[{\citenamefont{Gell-Mann et~al.}(1979)\citenamefont{Gell-Mann, Ramond,
  and Slansky}}]{GellMann:1980vs}
\bibinfo{author}{\bibfnamefont{M.}~\bibnamefont{Gell-Mann}},
  \bibinfo{author}{\bibfnamefont{P.}~\bibnamefont{Ramond}}, \bibnamefont{and}
  \bibinfo{author}{\bibfnamefont{R.}~\bibnamefont{Slansky}},
  \bibinfo{journal}{Conf. Proc.} \textbf{\bibinfo{volume}{C790927}},
  \bibinfo{pages}{315} (\bibinfo{year}{1979}), \eprint{1306.4669}.

\bibitem[{\citenamefont{Mohapatra and Senjanovic}(1980)}]{Mohapatra:1979ia}
\bibinfo{author}{\bibfnamefont{R.~N.} \bibnamefont{Mohapatra}}
  \bibnamefont{and}
  \bibinfo{author}{\bibfnamefont{G.}~\bibnamefont{Senjanovic}},
  \bibinfo{journal}{Phys. Rev. Lett.} \textbf{\bibinfo{volume}{44}},
  \bibinfo{pages}{912} (\bibinfo{year}{1980}).

\bibitem[{\citenamefont{Kibble et~al.}(1982)\citenamefont{Kibble, Lazarides,
  and Shafi}}]{Kibble:1982dd}
\bibinfo{author}{\bibfnamefont{T.~W.~B.} \bibnamefont{Kibble}},
  \bibinfo{author}{\bibfnamefont{G.}~\bibnamefont{Lazarides}},
  \bibnamefont{and} \bibinfo{author}{\bibfnamefont{Q.}~\bibnamefont{Shafi}},
  \bibinfo{journal}{Phys. Rev.} \textbf{\bibinfo{volume}{D26}},
  \bibinfo{pages}{435} (\bibinfo{year}{1982}).

\bibitem[{\citenamefont{Stern and Yajnik}(1986)}]{Stern:1985bg}
\bibinfo{author}{\bibfnamefont{A.}~\bibnamefont{Stern}} \bibnamefont{and}
  \bibinfo{author}{\bibfnamefont{U.~A.} \bibnamefont{Yajnik}},
  \bibinfo{journal}{Nucl. Phys.} \textbf{\bibinfo{volume}{B267}},
  \bibinfo{pages}{158} (\bibinfo{year}{1986}).

\bibitem[{\citenamefont{Bhattacharjee et~al.}(1992)\citenamefont{Bhattacharjee,
  Hill, and Schramm}}]{Pijush:1992}
\bibinfo{author}{\bibfnamefont{P.}~\bibnamefont{Bhattacharjee}},
  \bibinfo{author}{\bibfnamefont{C.~T.} \bibnamefont{Hill}}, \bibnamefont{and}
  \bibinfo{author}{\bibfnamefont{D.~N.} \bibnamefont{Schramm}},
  \bibinfo{journal}{Phys. Rev. Lett.} \textbf{\bibinfo{volume}{69}},
  \bibinfo{pages}{567} (\bibinfo{year}{1992}).

\bibitem[{\citenamefont{Davis and Jeannerot}(1995)}]{Davis:1995bx}
\bibinfo{author}{\bibfnamefont{A.-C.} \bibnamefont{Davis}} \bibnamefont{and}
  \bibinfo{author}{\bibfnamefont{R.}~\bibnamefont{Jeannerot}},
  \bibinfo{journal}{Phys. Rev.} \textbf{\bibinfo{volume}{D52}},
  \bibinfo{pages}{7220} (\bibinfo{year}{1995}), \eprint{hep-ph/9501275}.

\bibitem[{\citenamefont{Jeannerot et~al.}(2003)\citenamefont{Jeannerot, Rocher,
  and Sakellariadou}}]{Jeannerot:2003qv}
\bibinfo{author}{\bibfnamefont{R.}~\bibnamefont{Jeannerot}},
  \bibinfo{author}{\bibfnamefont{J.}~\bibnamefont{Rocher}}, \bibnamefont{and}
  \bibinfo{author}{\bibfnamefont{M.}~\bibnamefont{Sakellariadou}},
  \bibinfo{journal}{Phys. Rev.} \textbf{\bibinfo{volume}{D68}},
  \bibinfo{pages}{103514} (\bibinfo{year}{2003}), \eprint{hep-ph/0308134}.

\bibitem[{\citenamefont{Chakrabortty et~al.}(2017)\citenamefont{Chakrabortty,
  Maji, Mohanty, Patra, and Srivastava}}]{Chakrabortty:2017mgi}
\bibinfo{author}{\bibfnamefont{J.}~\bibnamefont{Chakrabortty}},
  \bibinfo{author}{\bibfnamefont{R.}~\bibnamefont{Maji}},
  \bibinfo{author}{\bibfnamefont{S.}~\bibnamefont{Mohanty}},
  \bibinfo{author}{\bibfnamefont{S.~K.} \bibnamefont{Patra}}, \bibnamefont{and}
  \bibinfo{author}{\bibfnamefont{T.}~\bibnamefont{Srivastava}}
  (\bibinfo{year}{2017}), \eprint{1711.11391}.

\bibitem[{\citenamefont{Aulakh and Girdhar}(2005)}]{Aulakh:2002zr}
\bibinfo{author}{\bibfnamefont{C.~S.} \bibnamefont{Aulakh}} \bibnamefont{and}
  \bibinfo{author}{\bibfnamefont{A.}~\bibnamefont{Girdhar}},
  \bibinfo{journal}{Int. J. Mod. Phys.} \textbf{\bibinfo{volume}{A20}},
  \bibinfo{pages}{865} (\bibinfo{year}{2005}), \eprint{hep-ph/0204097}.

\bibitem[{\citenamefont{Li et~al.}(2009)\citenamefont{Li, Wang, and
  Yang}}]{Li:2009sw}
\bibinfo{author}{\bibfnamefont{T.-j.} \bibnamefont{Li}},
  \bibinfo{author}{\bibfnamefont{F.}~\bibnamefont{Wang}}, \bibnamefont{and}
  \bibinfo{author}{\bibfnamefont{J.~M.} \bibnamefont{Yang}},
  \bibinfo{journal}{Nucl. Phys.} \textbf{\bibinfo{volume}{B820}},
  \bibinfo{pages}{534} (\bibinfo{year}{2009}), \eprint{0901.2161}.

\bibitem[{\citenamefont{Gogoladze et~al.}(2009)\citenamefont{Gogoladze, Rehman,
  and Shafi}}]{Gogoladze:2009bd}
\bibinfo{author}{\bibfnamefont{I.}~\bibnamefont{Gogoladze}},
  \bibinfo{author}{\bibfnamefont{M.~U.} \bibnamefont{Rehman}},
  \bibnamefont{and} \bibinfo{author}{\bibfnamefont{Q.}~\bibnamefont{Shafi}},
  \bibinfo{journal}{Phys. Rev.} \textbf{\bibinfo{volume}{D80}},
  \bibinfo{pages}{105002} (\bibinfo{year}{2009}), \eprint{0907.0728}.

\bibitem[{\citenamefont{Gogoladze et~al.}(2011)\citenamefont{Gogoladze, Khalid,
  Raza, and Shafi}}]{Gogoladze:2011db}
\bibinfo{author}{\bibfnamefont{I.}~\bibnamefont{Gogoladze}},
  \bibinfo{author}{\bibfnamefont{R.}~\bibnamefont{Khalid}},
  \bibinfo{author}{\bibfnamefont{S.}~\bibnamefont{Raza}}, \bibnamefont{and}
  \bibinfo{author}{\bibfnamefont{Q.}~\bibnamefont{Shafi}},
  \bibinfo{journal}{JHEP} \textbf{\bibinfo{volume}{06}}, \bibinfo{pages}{117}
  (\bibinfo{year}{2011}), \eprint{1102.0013}.

\bibitem[{\citenamefont{Lal~Awasthi and Parida}(2012)}]{LalAwasthi:2011aa}
\bibinfo{author}{\bibfnamefont{R.}~\bibnamefont{Lal~Awasthi}} \bibnamefont{and}
  \bibinfo{author}{\bibfnamefont{M.~K.} \bibnamefont{Parida}},
  \bibinfo{journal}{Phys. Rev.} \textbf{\bibinfo{volume}{D86}},
  \bibinfo{pages}{093004} (\bibinfo{year}{2012}), \eprint{1112.1826}.

\bibitem[{\citenamefont{Bertolini et~al.}(2012)\citenamefont{Bertolini,
  Di~Luzio, and Malinsky}}]{Bertolini:2012im}
\bibinfo{author}{\bibfnamefont{S.}~\bibnamefont{Bertolini}},
  \bibinfo{author}{\bibfnamefont{L.}~\bibnamefont{Di~Luzio}}, \bibnamefont{and}
  \bibinfo{author}{\bibfnamefont{M.}~\bibnamefont{Malinsky}},
  \bibinfo{journal}{Phys. Rev.} \textbf{\bibinfo{volume}{D85}},
  \bibinfo{pages}{095014} (\bibinfo{year}{2012}), \eprint{1202.0807}.

\bibitem[{\citenamefont{Mambrini et~al.}(2015)\citenamefont{Mambrini, Nagata,
  Olive, Quevillon, and Zheng}}]{Mambrini:2015vna}
\bibinfo{author}{\bibfnamefont{Y.}~\bibnamefont{Mambrini}},
  \bibinfo{author}{\bibfnamefont{N.}~\bibnamefont{Nagata}},
  \bibinfo{author}{\bibfnamefont{K.~A.} \bibnamefont{Olive}},
  \bibinfo{author}{\bibfnamefont{J.}~\bibnamefont{Quevillon}},
  \bibnamefont{and} \bibinfo{author}{\bibfnamefont{J.}~\bibnamefont{Zheng}},
  \bibinfo{journal}{Phys. Rev.} \textbf{\bibinfo{volume}{D91}},
  \bibinfo{pages}{095010} (\bibinfo{year}{2015}), \eprint{1502.06929}.

\bibitem[{\citenamefont{Blazek et~al.}(2002{\natexlab{a}})\citenamefont{Blazek,
  Dermisek, and Raby}}]{Blazek:2001sb}
\bibinfo{author}{\bibfnamefont{T.}~\bibnamefont{Blazek}},
  \bibinfo{author}{\bibfnamefont{R.}~\bibnamefont{Dermisek}}, \bibnamefont{and}
  \bibinfo{author}{\bibfnamefont{S.}~\bibnamefont{Raby}},
  \bibinfo{journal}{Phys. Rev. Lett.} \textbf{\bibinfo{volume}{88}},
  \bibinfo{pages}{111804} (\bibinfo{year}{2002}{\natexlab{a}}),
  \eprint{hep-ph/0107097}.

\bibitem[{\citenamefont{Blazek et~al.}(2002{\natexlab{b}})\citenamefont{Blazek,
  Dermisek, and Raby}}]{Blazek:2002ta}
\bibinfo{author}{\bibfnamefont{T.}~\bibnamefont{Blazek}},
  \bibinfo{author}{\bibfnamefont{R.}~\bibnamefont{Dermisek}}, \bibnamefont{and}
  \bibinfo{author}{\bibfnamefont{S.}~\bibnamefont{Raby}},
  \bibinfo{journal}{Phys. Rev.} \textbf{\bibinfo{volume}{D65}},
  \bibinfo{pages}{115004} (\bibinfo{year}{2002}{\natexlab{b}}),
  \eprint{hep-ph/0201081}.

\bibitem[{\citenamefont{Ross and Velasco-Sevilla}(2003)}]{Ross:2002fb}
\bibinfo{author}{\bibfnamefont{G.~G.} \bibnamefont{Ross}} \bibnamefont{and}
  \bibinfo{author}{\bibfnamefont{L.}~\bibnamefont{Velasco-Sevilla}},
  \bibinfo{journal}{Nucl. Phys.} \textbf{\bibinfo{volume}{B653}},
  \bibinfo{pages}{3} (\bibinfo{year}{2003}), \eprint{hep-ph/0208218}.

\bibitem[{\citenamefont{Dermisek et~al.}(2006)\citenamefont{Dermisek, Harada,
  and Raby}}]{Dermisek:2006dc}
\bibinfo{author}{\bibfnamefont{R.}~\bibnamefont{Dermisek}},
  \bibinfo{author}{\bibfnamefont{M.}~\bibnamefont{Harada}}, \bibnamefont{and}
  \bibinfo{author}{\bibfnamefont{S.}~\bibnamefont{Raby}},
  \bibinfo{journal}{Phys. Rev.} \textbf{\bibinfo{volume}{D74}},
  \bibinfo{pages}{035011} (\bibinfo{year}{2006}), \eprint{hep-ph/0606055}.

\bibitem[{\citenamefont{Aulakh and Mohapatra}(1983)}]{Aulakh:1982sw}
\bibinfo{author}{\bibfnamefont{C.~S.} \bibnamefont{Aulakh}} \bibnamefont{and}
  \bibinfo{author}{\bibfnamefont{R.~N.} \bibnamefont{Mohapatra}},
  \bibinfo{journal}{Phys. Rev.} \textbf{\bibinfo{volume}{D28}},
  \bibinfo{pages}{217} (\bibinfo{year}{1983}).

\bibitem[{\citenamefont{Babu and Mohapatra}(1993)}]{Babu:1992ia}
\bibinfo{author}{\bibfnamefont{K.~S.} \bibnamefont{Babu}} \bibnamefont{and}
  \bibinfo{author}{\bibfnamefont{R.~N.} \bibnamefont{Mohapatra}},
  \bibinfo{journal}{Phys. Rev. Lett.} \textbf{\bibinfo{volume}{70}},
  \bibinfo{pages}{2845} (\bibinfo{year}{1993}), \eprint{hep-ph/9209215}.

\bibitem[{\citenamefont{Aulakh et~al.}(2004)\citenamefont{Aulakh, Bajc, Melfo,
  Senjanovic, and Vissani}}]{Aulakh:2003kg}
\bibinfo{author}{\bibfnamefont{C.~S.} \bibnamefont{Aulakh}},
  \bibinfo{author}{\bibfnamefont{B.}~\bibnamefont{Bajc}},
  \bibinfo{author}{\bibfnamefont{A.}~\bibnamefont{Melfo}},
  \bibinfo{author}{\bibfnamefont{G.}~\bibnamefont{Senjanovic}},
  \bibnamefont{and} \bibinfo{author}{\bibfnamefont{F.}~\bibnamefont{Vissani}},
  \bibinfo{journal}{Phys. Lett.} \textbf{\bibinfo{volume}{B588}},
  \bibinfo{pages}{196} (\bibinfo{year}{2004}), \eprint{hep-ph/0306242}.

\bibitem[{\citenamefont{Bajc et~al.}(2004{\natexlab{a}})\citenamefont{Bajc,
  Melfo, Senjanovic, and Vissani}}]{Bajc:2004xe}
\bibinfo{author}{\bibfnamefont{B.}~\bibnamefont{Bajc}},
  \bibinfo{author}{\bibfnamefont{A.}~\bibnamefont{Melfo}},
  \bibinfo{author}{\bibfnamefont{G.}~\bibnamefont{Senjanovic}},
  \bibnamefont{and} \bibinfo{author}{\bibfnamefont{F.}~\bibnamefont{Vissani}},
  \bibinfo{journal}{Phys. Rev.} \textbf{\bibinfo{volume}{D70}},
  \bibinfo{pages}{035007} (\bibinfo{year}{2004}{\natexlab{a}}),
  \eprint{hep-ph/0402122}.

\bibitem[{\citenamefont{Bajc et~al.}(2004{\natexlab{b}})\citenamefont{Bajc,
  Senjanovic, and Vissani}}]{Bajc:2004fj}
\bibinfo{author}{\bibfnamefont{B.}~\bibnamefont{Bajc}},
  \bibinfo{author}{\bibfnamefont{G.}~\bibnamefont{Senjanovic}},
  \bibnamefont{and} \bibinfo{author}{\bibfnamefont{F.}~\bibnamefont{Vissani}},
  \bibinfo{journal}{Phys. Rev.} \textbf{\bibinfo{volume}{D70}},
  \bibinfo{pages}{093002} (\bibinfo{year}{2004}{\natexlab{b}}),
  \eprint{hep-ph/0402140}.

\bibitem[{\citenamefont{Aulakh and Garg}(2012{\natexlab{a}})}]{Aulakh:2008sn}
\bibinfo{author}{\bibfnamefont{C.~S.} \bibnamefont{Aulakh}} \bibnamefont{and}
  \bibinfo{author}{\bibfnamefont{S.~K.} \bibnamefont{Garg}},
  \bibinfo{journal}{Nucl. Phys.} \textbf{\bibinfo{volume}{B857}},
  \bibinfo{pages}{101} (\bibinfo{year}{2012}{\natexlab{a}}),
  \eprint{0807.0917}.

\bibitem[{\citenamefont{Aulakh and Garg}(2012{\natexlab{b}})}]{Aulakh:2012st}
\bibinfo{author}{\bibfnamefont{C.~S.} \bibnamefont{Aulakh}} \bibnamefont{and}
  \bibinfo{author}{\bibfnamefont{I.}~\bibnamefont{Garg}},
  \bibinfo{journal}{Phys. Rev.} \textbf{\bibinfo{volume}{D86}},
  \bibinfo{pages}{065001} (\bibinfo{year}{2012}{\natexlab{b}}),
  \eprint{1201.0519}.

\bibitem[{\citenamefont{Garg and Mohanty}(2015)}]{Garg:2015mra}
\bibinfo{author}{\bibfnamefont{I.}~\bibnamefont{Garg}} \bibnamefont{and}
  \bibinfo{author}{\bibfnamefont{S.}~\bibnamefont{Mohanty}},
  \bibinfo{journal}{Phys. Lett.} \textbf{\bibinfo{volume}{B751}},
  \bibinfo{pages}{7} (\bibinfo{year}{2015}), \eprint{1504.07725}.

\bibitem[{\citenamefont{Aulakh et~al.}(1997)\citenamefont{Aulakh, Benakli, and
  Senjanovic}}]{Aulakh:1997ba}
\bibinfo{author}{\bibfnamefont{C.~S.} \bibnamefont{Aulakh}},
  \bibinfo{author}{\bibfnamefont{K.}~\bibnamefont{Benakli}}, \bibnamefont{and}
  \bibinfo{author}{\bibfnamefont{G.}~\bibnamefont{Senjanovic}},
  \bibinfo{journal}{Phys. Rev. Lett.} \textbf{\bibinfo{volume}{79}},
  \bibinfo{pages}{2188} (\bibinfo{year}{1997}), \eprint{hep-ph/9703434}.

\bibitem[{\citenamefont{Mishra and Yajnik}(2010)}]{Mishra:2009mk}
\bibinfo{author}{\bibfnamefont{S.}~\bibnamefont{Mishra}} \bibnamefont{and}
  \bibinfo{author}{\bibfnamefont{U.~A.} \bibnamefont{Yajnik}},
  \bibinfo{journal}{Phys. Rev.} \textbf{\bibinfo{volume}{D81}},
  \bibinfo{pages}{045010} (\bibinfo{year}{2010}), \eprint{0911.1578}.

\bibitem[{\citenamefont{Mishra et~al.}(2009)\citenamefont{Mishra, Yajnik, and
  Sarkar}}]{Mishra:2008be}
\bibinfo{author}{\bibfnamefont{S.}~\bibnamefont{Mishra}},
  \bibinfo{author}{\bibfnamefont{U.~A.} \bibnamefont{Yajnik}},
  \bibnamefont{and} \bibinfo{author}{\bibfnamefont{A.}~\bibnamefont{Sarkar}},
  \bibinfo{journal}{Phys. Rev.} \textbf{\bibinfo{volume}{D79}},
  \bibinfo{pages}{065038} (\bibinfo{year}{2009}), \eprint{0812.0868}.

\bibitem[{\citenamefont{Sarkar et~al.}(2008)\citenamefont{Sarkar, Abhishek, and
  Yajnik}}]{Sarkar:2007er}
\bibinfo{author}{\bibfnamefont{A.}~\bibnamefont{Sarkar}},
  \bibinfo{author}{\bibnamefont{Abhishek}}, \bibnamefont{and}
  \bibinfo{author}{\bibfnamefont{U.~A.} \bibnamefont{Yajnik}},
  \bibinfo{journal}{Nucl. Phys.} \textbf{\bibinfo{volume}{B800}},
  \bibinfo{pages}{253} (\bibinfo{year}{2008}), \eprint{0710.5410}.

\bibitem[{\citenamefont{Linde}(1983)}]{Linde:1983gd}
\bibinfo{author}{\bibfnamefont{A.~D.} \bibnamefont{Linde}},
  \bibinfo{journal}{Phys. Lett.} \textbf{\bibinfo{volume}{129B}},
  \bibinfo{pages}{177} (\bibinfo{year}{1983}).

\bibitem[{\citenamefont{Martin and Ringeval}(2010)}]{Martin:2010kz}
\bibinfo{author}{\bibfnamefont{J.}~\bibnamefont{Martin}} \bibnamefont{and}
  \bibinfo{author}{\bibfnamefont{C.}~\bibnamefont{Ringeval}},
  \bibinfo{journal}{Phys. Rev.} \textbf{\bibinfo{volume}{D82}},
  \bibinfo{pages}{023511} (\bibinfo{year}{2010}), \eprint{1004.5525}.

\bibitem[{\citenamefont{Dai et~al.}(2014)\citenamefont{Dai, Kamionkowski, and
  Wang}}]{Dai:2014jja}
\bibinfo{author}{\bibfnamefont{L.}~\bibnamefont{Dai}},
  \bibinfo{author}{\bibfnamefont{M.}~\bibnamefont{Kamionkowski}},
  \bibnamefont{and} \bibinfo{author}{\bibfnamefont{J.}~\bibnamefont{Wang}},
  \bibinfo{journal}{Phys. Rev. Lett.} \textbf{\bibinfo{volume}{113}},
  \bibinfo{pages}{041302} (\bibinfo{year}{2014}), \eprint{1404.6704}.

\bibitem[{\citenamefont{Cook et~al.}(2015)\citenamefont{Cook, Dimastrogiovanni,
  Easson, and Krauss}}]{Cook:2015vqa}
\bibinfo{author}{\bibfnamefont{J.~L.} \bibnamefont{Cook}},
  \bibinfo{author}{\bibfnamefont{E.}~\bibnamefont{Dimastrogiovanni}},
  \bibinfo{author}{\bibfnamefont{D.~A.} \bibnamefont{Easson}},
  \bibnamefont{and} \bibinfo{author}{\bibfnamefont{L.~M.}
  \bibnamefont{Krauss}}, \bibinfo{journal}{JCAP}
  \textbf{\bibinfo{volume}{1504}}, \bibinfo{pages}{047} (\bibinfo{year}{2015}),
  \eprint{1502.04673}.

\bibitem[{\citenamefont{Goswami and Yajnik}(2018)}]{Rajesh:2018}
\bibinfo{author}{\bibfnamefont{R.}~\bibnamefont{Goswami}} \bibnamefont{and}
  \bibinfo{author}{\bibfnamefont{U.~A.} \bibnamefont{Yajnik}},
  \bibinfo{journal}{in preparation}  (\bibinfo{year}{2018}).

\bibitem[{\citenamefont{Smoot et~al.}(1992)}]{Smoot:1992td}
\bibinfo{author}{\bibfnamefont{G.~F.} \bibnamefont{Smoot}} \bibnamefont{et~al.}
  (\bibinfo{collaboration}{COBE}), \bibinfo{journal}{Astrophys. J.}
  \textbf{\bibinfo{volume}{396}}, \bibinfo{pages}{L1} (\bibinfo{year}{1992}).

\bibitem[{\citenamefont{Bennett et~al.}(2013)}]{Bennett:2012zja}
\bibinfo{author}{\bibfnamefont{C.~L.} \bibnamefont{Bennett}}
  \bibnamefont{et~al.} (\bibinfo{collaboration}{WMAP}),
  \bibinfo{journal}{Astrophys. J. Suppl.} \textbf{\bibinfo{volume}{208}},
  \bibinfo{pages}{20} (\bibinfo{year}{2013}), \eprint{1212.5225}.

\bibitem[{\citenamefont{Bouhmadi-Lopez
  et~al.}(2013)\citenamefont{Bouhmadi-Lopez, Chen, Huang, and
  Lin}}]{BouhmadiLopez:2012by}
\bibinfo{author}{\bibfnamefont{M.}~\bibnamefont{Bouhmadi-Lopez}},
  \bibinfo{author}{\bibfnamefont{P.}~\bibnamefont{Chen}},
  \bibinfo{author}{\bibfnamefont{Y.-C.} \bibnamefont{Huang}}, \bibnamefont{and}
  \bibinfo{author}{\bibfnamefont{Y.-H.} \bibnamefont{Lin}},
  \bibinfo{journal}{Phys. Rev.} \textbf{\bibinfo{volume}{D87}},
  \bibinfo{pages}{103513} (\bibinfo{year}{2013}), \eprint{1212.2641}.

\bibitem[{\citenamefont{Kolb and Turner}(1990)}]{Kolb:1990vq}
\bibinfo{author}{\bibfnamefont{E.~W.} \bibnamefont{Kolb}} \bibnamefont{and}
  \bibinfo{author}{\bibfnamefont{M.~S.} \bibnamefont{Turner}},
  \bibinfo{journal}{Front. Phys.} \textbf{\bibinfo{volume}{69}},
  \bibinfo{pages}{1} (\bibinfo{year}{1990}).

\bibitem[{\citenamefont{Durrer et~al.}(2002)\citenamefont{Durrer, Kunz, and
  Melchiorri}}]{Durrer:2001cg}
\bibinfo{author}{\bibfnamefont{R.}~\bibnamefont{Durrer}},
  \bibinfo{author}{\bibfnamefont{M.}~\bibnamefont{Kunz}}, \bibnamefont{and}
  \bibinfo{author}{\bibfnamefont{A.}~\bibnamefont{Melchiorri}},
  \bibinfo{journal}{Phys. Rept.} \textbf{\bibinfo{volume}{364}},
  \bibinfo{pages}{1} (\bibinfo{year}{2002}), \eprint{astro-ph/0110348}.

\bibitem[{\citenamefont{Urrestilla et~al.}(2008)\citenamefont{Urrestilla,
  Bevis, Hindmarsh, Kunz, and Liddle}}]{Urrestilla:2007sf}
\bibinfo{author}{\bibfnamefont{J.}~\bibnamefont{Urrestilla}},
  \bibinfo{author}{\bibfnamefont{N.}~\bibnamefont{Bevis}},
  \bibinfo{author}{\bibfnamefont{M.}~\bibnamefont{Hindmarsh}},
  \bibinfo{author}{\bibfnamefont{M.}~\bibnamefont{Kunz}}, \bibnamefont{and}
  \bibinfo{author}{\bibfnamefont{A.~R.} \bibnamefont{Liddle}},
  \bibinfo{journal}{JCAP} \textbf{\bibinfo{volume}{0807}}, \bibinfo{pages}{010}
  (\bibinfo{year}{2008}), \eprint{0711.1842}.

\bibitem[{\citenamefont{Ringeval}(2010)}]{Ringeval:2010ca}
\bibinfo{author}{\bibfnamefont{C.}~\bibnamefont{Ringeval}},
  \bibinfo{journal}{Adv. Astron.} \textbf{\bibinfo{volume}{2010}},
  \bibinfo{pages}{380507} (\bibinfo{year}{2010}), \eprint{1005.4842}.

\bibitem[{\citenamefont{Ade et~al.}(2014)}]{Ade:2013xla}
\bibinfo{author}{\bibfnamefont{P.~A.~R.} \bibnamefont{Ade}}
  \bibnamefont{et~al.} (\bibinfo{collaboration}{Planck}),
  \bibinfo{journal}{Astron. Astrophys.} \textbf{\bibinfo{volume}{571}},
  \bibinfo{pages}{A25} (\bibinfo{year}{2014}), \eprint{1303.5085}.

\end{thebibliography}
\end{document}